# Power Network SCADA Quantum Communications: A Comparison of BB84, B92, E91, and SGS04 Quantum Key Distribution Protocols

Hillol Biswas* Kyriakos E. Zoiros

Department of Electrical and Computer Engineering,

Democritus University of Thrace,

Xanthi, Greece

## Abstract:

The current state, emerging trends, and practical challenges of optical fiber-based power network SCADA quantum communication must be addressed to fully utilise the technological platform's potential in real-world power system SCADA communications involving massive volumes of real-time data, as well as in managing, encoding, and applications such as quantum cryptography. Quantum key distribution (QKD) is an essential part of the cybersecurity paradigm for quantum communication. Even though quantum computing with individual circuits yields probabilistic outcomes for the problem at hand, real-world datasets are complex and challenging to handle, even with telemetry. When using the cybersecurity triad of availability, confidentiality, and integrity (CIA) in reverse order (AIC), availability is given priority in electric power networks. This research assesses the use of the BB84, E91, B92, and SARG04 cryptographic protocols by applying them to large, multivariate power-system SCADA datasets and comparing the outcomes. By leveraging the variety of QKD protocols available with quantum electronics hardware, this simulation work provides a promising avenue for developing implementable frameworks for SCADA/PMU networks in power systems.

## 1.0 Introduction

*Power system communication infrastructure entails* the complexity of current communication in power networks, which requires installing optical fiber ground wire, typically 24-core to 96-core, over long-distance transmission lines, either overhead or underground. The OPGW terminates in the substation, where it is laid through a cable trench within the substation yard and control room. In the control room, Ethernet connections are made through the substation automation system communication panel and connect to other panels, such as transformers, bus couplers, and others, depending on the substation size and complexity. The complexity further increases when adhering to the respective monitoring, control, protection, and communication industry standards, usually based on ITU's various series.

*SCADA and substation automation are integral parts of the ecosystem.* Supervisory Control and Data Acquisition is referred to as SCADA. As the name suggests, it focuses on the supervisory level rather than providing full control. Because of this, it is a software-only package installed on top of hardware it interfaces with, usually via PLCs, IEDs, or other commercial hardware modules. Most industrial processes, including those in steel production, nuclear and conventional electricity generation and distribution, chemistry, and some experimental facilities like nuclear fusion, use SCADA systems [1].

At reduced maintenance resources and costs, automated substations can provide the data needed to keep the customer's power supply constant. The integration of intelligent electrical devices (such as circuit breakers, transformers, and relays) with the ability to track their operation is known as substation automation. Circuit breakers, for instance, can detect appropriate maintenance diagnostics and assess their contact resistance. Microprocessor-based relays, circuit breakers, transformers, and motor-operated air switches make up automated substations. These components are all under the control of a remotely accessible graphical user interface unit [2].

The SCADA framework comprises both software and hardware. The software consists of "Human Machine Interface (HMI)," a central database (Historian), and other user software, while the hardware consists of "Remote Terminal Units (RTU)," "Master Terminal Unit (MTU)," actuators, and sensors [3]. These programs offer a means of communication between software and hardware. Actuators and sensors are connected to the physical environment, which is then connected to RTUs. To monitor and operate the SCADA system, RTUs collect sensor data and transmit telemetry to the MTU [4]. It has components and communication protocols that frame its architecture; however, Power System SCADA and its protocols are unique to power system operations in electricity utility networks. Since the introduction of the IEC 61850 series over the last two decades, utilities worldwide have adopted this interface for power system monitoring, control, and protection. It maps to the OSI (Open Systems Interconnection) Model, a set of guidelines that describes how various computer systems interact across a network. The International Organization for Standardization (ISO) created the OSI Model, which is a 7-layer system. Each of the seven layers that make up the OSI Model has distinct roles and duties. This tiered approach enables collaboration among many devices and technologies. The OSI Model provides a clear framework for handling network issues and data transfer. Many utility domains use the OSI Model as a guide to comprehend how network systems work [5].

*Cybersecurity in smart grids is also imperative being one of the building blocks,* as they entail bidirectional power and data flows; it is a key component of power system operation for protecting the grid, depending on its size and complexity. The complexity is multifaceted, and, given the growing cybersecurity landscape in this critical infrastructure domain, emerging threats and vulnerabilities are paramount.

*In line with Quantum Communication Motivation and Growth,* as quantum computing and related technologies have attracted significant research interest across academia and industry, quantum communication in power system operation has also attracted growing interest worldwide. Many utilities are considering implementing it experimentally, while research on quantum communication through optical fibre has also been expanding.

*In the Real-World Implementations and Projects, viz.,* the Danish Grid, for instance, has installed optical fiber ground wire and optical fiber cable-based quantum key transmission for the electricity power system in the substation at Funen from the Frøslev converter station [6]. Research and application-oriented efforts on the use of quantum encryption in power systems and the utility sector are ongoing. In early February 2019, the fiber network controlled by EPB in Chattanooga, Tennessee, implemented and demonstrated a trusted node quantum key distribution (QKD) system that combines two QKD techniques [7], [8].

LANL developed hardware and software to enhance security and reduce costs for quantum communication nodes that deliver QKD. Oak Ridge National Laboratory's (ORNL) "Quantum Physics Secured Communications for the Energy Sector" (Q-Sens) project. Q-Sens addresses the cost and distance limitations of QKD by providing new quantum protocols for data authentication [9].

To secure Geneva's smart grid networks, QKD technology was examined and evaluated in an operational environment. It used both conventional channels in grid networks and a direct connection with a 3.4 km-long dark fiber specifically constructed for QKD [10].

*Pertaining to Advanced Research Directions,* **a**lthough the study acknowledges insufficient data, which is crucial for any real-world power network multidisciplinary domain, a simulated attack centred on the load frequency control (LFC) and using a Quantum Machine Learning (QML) technique has been described [11]. With ever-increasing data availability and processing, smart grid 2.0 is expected to facilitate quantum-variants-based steps toward a projected carbon-neutral goal around 2050 [12].

*In QKD Protocol Details and Applications,* the IKE algorithm is used in the sorting phase. Similar to the SARG04 protocol, the proposed method is more resilient to PNS attacks [13]. In QKD, once the error rate is low enough and enough photons have been detected, a shared private key known only to the sender and the recipient can be generated. It explains how this cutting-edge technology and its various modalities could benefit the essential infrastructure of dams and hydropower plants [14].

Thus, from communication to electrical power systems operations for protecting the grid, it is imperative that any identified application areas in the quantum regime ideally require a broad overview of domain knowledge, as they are intersectional across domains. As cyber attackers' growing sophistication becomes apparent, it is likely surprising that future cybersecurity breaches will not, in turn, entail quantum attacks.

The paper is organized into the following sections: 2.0, Background; 3.0, Methods; Simulation; and Experimental Studies. The 4.0, results section presents the findings, followed by the discussion, and the 5. 0, the conclusion section wraps up the paper.

# 2.0 Background

Power networks' SCADA communication and cybersecurity are interwoven in the operational regime to transmit power from one node to another node. In the growing research and implementation landscape of quantum communication for secure communication using quantum-mechanical laws of physics in national critical infrastructures and to protect future power grids, it is imperative that QKD implementation be robust and realistic, leveraging existing protocols, albeit customized to real-world scenarios.

This section comprises sub-sections of optical fiber-based communication infrastructure, operational technology, and SCADA security, intrusion detection and prevention systems (IDPS), foundations of quantum communication, optical quantum communication and QKD systems, quantum communications in SCADA and smart grids, & quantum algorithms and cryptographic implementations, as given below:

## 2.1 Optical Fiber-Based Communication Infrastructure

Communication channels for modern networks may be broadly categorized into free-space links (e.g., satellite communication) and optical fiber-based systems. Given the widespread deployment of optical fiber in existing telecommunication infrastructure, particularly in Internet backbones, fiber-optic links remain the most practical and scalable medium for high-capacity data transmission. Various fiber types—including single-mode fibers (SMFs), multi-mode fibers (MMFs), and multicore fibers (MCFs)—enable the propagation of high-dimensional quantum states, with the choice of fiber depending on application-specific constraints. For instance, MCFs are particularly suitable for data center environments where space efficiency and high connectivity density are critical.

In power system communication, optical fiber ground wire (OPGW) typically employs ITU-T G.652 compliant single-mode fibers [15]. The ITU-T G.652.D standard specifies key physical and transmission parameters such as mode field diameter, cladding characteristics, attenuation, chromatic dispersion, and polarization mode dispersion. Additionally, ITU-T L.151 provides detailed recommendations for the installation and deployment of OPGW cables, including handling procedures, environmental considerations, and jointing techniques [16]. These standards ensure reliability and robustness in harsh outdoor transmission environments.

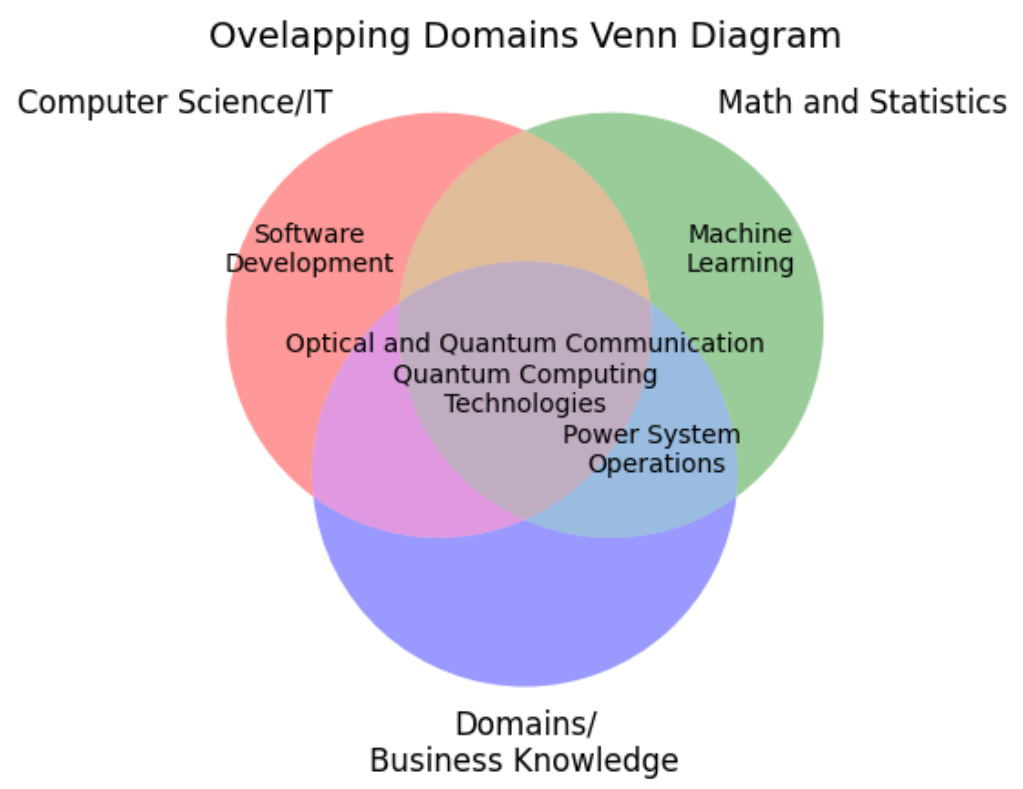

*Fig. 1: Overlapping cum intersection domains of application*

As the topic is an intersection of a few relevant domains [17] Fig. 1 illustrates the area of intersection of software/IT knowledge, artificial intelligence cum machine learning, and the power system operations, having sub-domains of optical and quantum communication, cyber security in operational technology (OT), quantum computing, and technologies of quantum hardware, i.e., quantum electronics sub-domain.

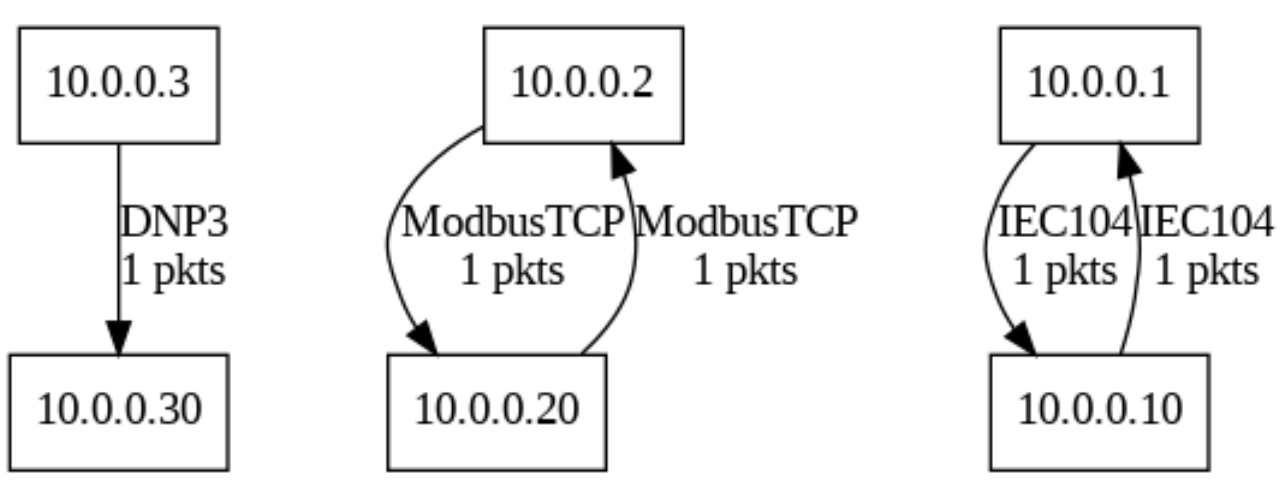

*Fig. 2: SCADA three protocol sections*

Fig. 2 depicts a typical master-to-RTU polling exchange with a single observed packet, shown by the unidirectional DNP3 communication flow from host 10.0.0.3 to 10.0.0.30 on the left side of the diagram. A typical client-server request-response transaction with one packet in each direction is depicted in the middle section, showing a bidirectional Modbus TCP connection between 10.0.0.2 and 10.0.0.20. A bidirectional IEC 60870-5-104 communication between 10.0.0.1 and 10.0.0.10 is shown on the right, representing a standard SCADA master–substation exchange in which a single packet is sent and received. When taken as a whole, the figure shows how several industrial control protocols can operate simultaneously within a single network segment. From cybersecurity paradigm, each protocol being active simultaneously have own weakness though.

10.0.0.1 ◄ IEC104 | TID=100, CoT=3 | 20B | 1.001234 ► 10.0.0.10

IEC104 | TID=100, CoT=7 | 18B | 1.101500

*Fig. 3: IEC 104_sequence_graph*

An IEC 60870-5-104 communication exchange between 10.0.0.1 and 10.0.0.10 is depicted in the Fig. 3. A command or interrogation-related transmission is indicated by the message's Type Identifier (TID) = 100, Cause of Transmission (CoT) = 3, payload size of 20 bytes, and timestamp 1.001234. With TID = 100 and CoT = 7 (18 bytes, timestamp 1.101500), a matching IEC 104 frame indicates a follow-up response or activation confirmation. When combined, the exchange exemplifies a standard IEC 104 request-response exchange in a communication link between a SCADA master and substation.

In a substation automation system (SAS), communication protocols are mapped to power system operations for monitoring, control, and protection, as shown in the IEC 61850 series, which are interfaces rather than a distinct protocol or set of protocols.

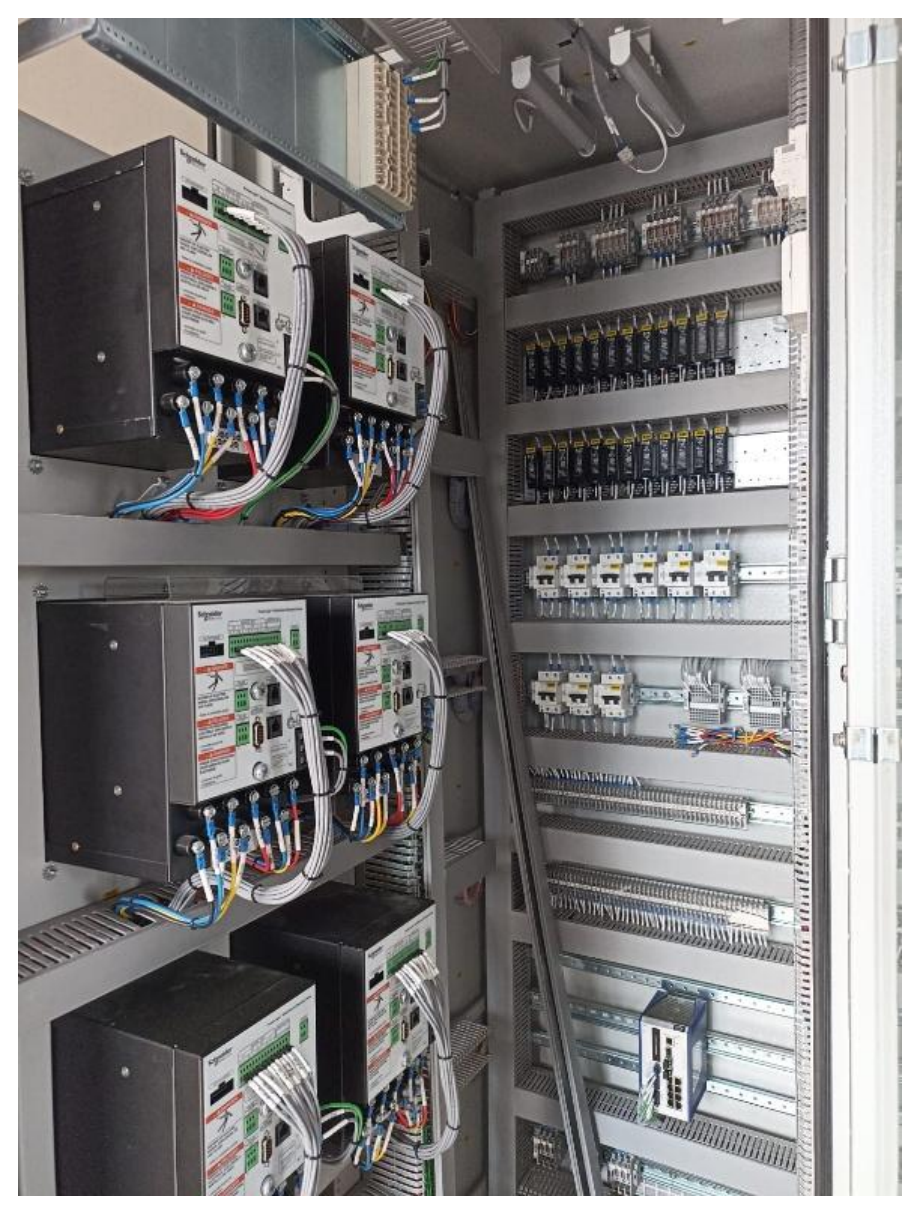

*Fig. 4: A typical SCADA Substation Automation System Panel Inside*

A typical SCADA substation panel inside is dedicated to SCADA communication and control. The SCADA communication and control panel's internal layout is depicted in Fig. 4. It includes several industrial control modules, communication devices, protection relays, and terminal blocks installed on DIN rails. Relays, terminal strips, and circuit protection components are arranged in the right portion, while power/communication modules with Ethernet and field wiring connections are in the left. A standard industrial control cabinet design that facilitates remote monitoring, control signalling, and protocol-based communication (e.g., IEC 104, Modbus, DNP3) between field devices and the control centre is shown by the structured cable routing and segregation. As seen various Protection Relays (IEDs,

Network Switch, Auxiliary Relays, MCBs (Miniature Circuit Breakers), Terminal Blocks, Power Supply Units are visible within the panel.

Together, the cabinet image depicts the actual electrical equipment that creates and handles these connections, whereas the previous network diagrams show the logical SCADA protocol exchanges (IEC 104, Modbus TCP, DNP3). RTUs, PLCs, communication processors, and power conditioning units are among the hardware modules that make up the panel's electronics layer, which connects field signals to IP-based industrial protocols. As a result, the network diagrams' protocol-level traffic clearly represents the electronic control and communication hardware located inside the SCADA panel.

## 2.2 Operational Technology (OT) and SCADA Security

Power systems rely heavily on Supervisory Control and Data Acquisition (SCADA) systems, which are increasingly exposed to cybersecurity threats. The Aurora experiment demonstrated that malicious manipulation of SCADA control signals can lead to severe physical damage to power infrastructure [18]. Unlike traditional information technology (IT) systems, operational technology (OT) prioritizes availability over confidentiality, reversing the conventional CIA (Confidentiality, Integrity, Availability) triad.

The cybersecurity landscape for power systems has evolved significantly, with numerous incidents reported globally [20]. While SCADA connectivity enhances scalability and remote operation, it simultaneously introduces vulnerabilities by exposing otherwise isolated systems to cyber threats [19]. Furthermore, SCADA architectures rely on diverse communication protocols such as IEC 60870-101/104, DNP-3, Fieldbus, and Profibus, contributing to a complex and heterogeneous security environment [20], [4], [21].

## 2.3 Intrusion Detection and Prevention Systems (IDPS)

Intrusion Detection and Prevention Systems (IDPS) play a crucial role in safeguarding networked infrastructures. Intrusion detection involves monitoring system or network activities for signs of security violations, while intrusion prevention extends this capability by actively mitigating detected threats. Due to overlapping functionalities, the combined term IDPS is commonly used [22].

IDPS mechanisms typically analyze traffic across multiple layers of the TCP/IP stack, including application, transport, network, and data link layers. Detection techniques are broadly classified into signature-based, anomaly-based, and stateful protocol analysis approaches [22]. Modern systems often employ hybrid techniques to enhance detection accuracy.

In SCADA environments, network monitoring tools such as Wireshark [23], Suricata [24], and Snort [25] are widely used for packet inspection and real-time situational awareness. These tools support protocol dissection and traffic logging, forming an essential component of cybersecurity monitoring frameworks.

## 2.4 Foundations of Quantum Communication

Quantum communication leverages fundamental principles of quantum mechanics to ensure secure information transfer. The Heisenberg Uncertainty Principle [26] implies that any measurement disturbs the quantum state, enabling the detection of eavesdropping attempts. Similarly, the No-Cloning Theorem [27] prohibits the exact replication of arbitrary quantum states, forming the basis for secure quantum communication protocols.

From an information-theoretic perspective, Holevo's theorem establishes limits on the amount of classical information that can be transmitted using quantum systems [38]. Despite these limits, quantum communication offers significant advantages over classical approaches through superposition and entanglement. These properties enable more efficient encoding of correlations and can reduce communication complexity, allowing certain computational tasks to be performed with exponentially fewer resources than classical methods [28].

## 2.5 Optical Quantum Communication and QKD Systems

Optical fibers are particularly well-suited for quantum communication due to their compatibility with existing telecom infrastructure and low-loss transmission at wavelengths around 1550 nm [29], [30]. A typical fiber-based quantum key distribution (QKD) system consists of a low-loss channel (~0.2 dB/km), quantum sources (e.g., single photons or weak coherent pulses), encoding schemes (polarization, phase, time-bin, or quadrature), and detectors such as InGaAs avalanche photodiodes or superconducting nanowire single-photon detectors. Classical post-processing steps—including reconciliation and privacy amplification—are used to generate secure keys [29].

Several QKD protocols have been developed. Discrete-variable QKD (DV-QKD), including BB84 and its decoy-state variants, is the most mature approach [31]. Continuous-variable QKD (CV-QKD) utilizes homodyne or heterodyne detection and integrates well with standard telecom components [32], [33]. Entanglement-based protocols, such as BBM92/B82, provide enhanced security guarantees [29], while advanced schemes like measurement-device-independent (MDI) QKD and twin-field QKD improve long-distance performance and resilience to detector-side attacks [34].

Despite significant progress, QKD performance is constrained by channel attenuation, detector noise, and finite-size effects [3], [4]. Practical deployments have demonstrated secure communication over metropolitan and intercity distances, with field trials exceeding a few hundred km in optical fiber networks [4], [35]. Trusted-node architectures are commonly employed to extend operational range.

Quantum repeaters, which utilize entanglement swapping and quantum memories, represent a long-term solution for overcoming distance limitations, although they remain under active development [36], [37], [38]. Practical challenges such as Raman noise, coexistence with classical signals, and system cost are addressed through filtering techniques, wavelength management, and photonic integration [39]. Typical fiber losses range from 0.17 to 0.25 dB/km, with secure key rates varying from kilobits per second in metropolitan networks to below 1 bit per second over long distances [40], [8].

Standardization efforts are ongoing, with organizations such as the ITU actively promoting QKD and post-quantum cryptography (PQC) as complementary approaches [41]. Notable deployments include the Micius satellite and the Beijing–Shanghai QKD backbone, highlighting the transition of QKD from research to real-world applications, despite challenges related to cost and scalability [35].

## 2.6 Quantum Communication in SCADA and Smart Grids

The integration of quantum communication into power system networks has gained increasing attention. QKD-based secure communication frameworks have been proposed for SCADA systems and smart grid infrastructures [21]. These approaches aim to enhance resilience against both classical and quantum-enabled cyber threats.

Recent works have explored hybrid architectures combining QKD with emerging technologies. For instance, QPUF-based device authentication integrated with QKD has been evaluated using quantum simulators from IBM and Rigetti [42]. Similarly, blockchain-based frameworks such as the Self-Defensive Post-Quantum Blockchain Architecture (SD-PQBA) have been proposed to secure SCADA systems in smart city environments [43].

Ensuring cybersecurity in industrial control systems (ICS) remains a critical challenge, necessitating multi-layered defense strategies and continuous evaluation of emerging threats [44], [45].

## 2.7 Quantum Algorithms and Cryptographic Implications

Quantum computing poses significant challenges to classical cryptographic systems. Algorithms such as Shor's algorithm threaten the security of widely used public-key schemes like RSA, while Grover's algorithm reduces the effective security of symmetric encryption methods such as AES [46], [47]. These vulnerabilities have motivated the exploration of quantum-resistant and quantum-based cryptographic solutions.

Quantum SIGNCRYPTION techniques have been proposed for SCADA systems, leveraging quantum principles for secure communication [48], albeit generic use for SCADA application domains. Additionally, the no-cloning theorem ensures that any attempt to duplicate quantum information introduces detectable disturbances, reinforcing the security of quantum communication systems. Experimental studies have demonstrated optimal quantum cloning under physical constraints, particularly for low-dimensional photonic states, although scalability remains a challenge [49].

# 3.0 Methods, Simulation, and Experimental Studies

Simulation and experimental validation are vital for advancing quantum-secured communication systems. Frameworks integrating QKD with quantum hardware and simulators, such as those provided by IBM and Rigetti, have been used to evaluate system performance and feasibility [42].

In parallel, classical cybersecurity techniques continue to evolve. Packet analysis tools, including Wireshark and TCPDump, are widely used for network monitoring and are increasingly combined with artificial intelligence and machine learning methods for enhanced intrusion detection [50]. Experimental studies on transport layer security (TLS) have also investigated the use of RSA and Elliptic Curve Digital Signature Algorithm (ECDSA) for secure certificate signing in SCADA communication systems [51].

Overall, the literature demonstrates a clear transition from classical communication and cybersecurity mechanisms to quantum-enhanced secure communication frameworks. Optical fiber-based QKD has emerged as the most mature approach for short- to medium-distance deployment, while integration with SCADA and smart grid systems represents a promising direction for securing critical infrastructure in the quantum era. Therefore, the obvious question surfaces: which QKD to select? Are the available ones equally applicable in this specific domain application area?

## 3.1 Contribution and Novelty:

The main contributions of this paper highlight the general applicability of four different QKDs across power system SCADA protocols, open-source testbed datasets, and real-world telemetry scenarios. While adopting QKD based on known protocols customized for the multivariate datasets, the paper's novelty lies in comparing four QKD protocols, viz., BB84, B92, E91, and SGS04, to investigate the generic applicability and evaluate the scope of selection of a specific protocol from the cybersecurity regime. Nonetheless, this is the first approach to compare four QKDs in power systems SCADA datasets.

This section is split into the following sub-sections, viz. The paper presents QKD Protocols, including BB84, B92, E91, and SGS04, along with approach and follow-up simulations. The contribution and novelty sub-section highlights the paper's unique approach and implementation.

## 3.2 QKD Protocols

Using quantum-mechanical principles, OKD demonstrates secure communication methods. The core idea is that any eavesdropping attempts disturb the quantum states and are thus detectable. Table 1 compares the four core QKD protocols, their year of introduction, the corresponding basis states, and the security basis.

*Table 1: Comparison Table of Quantum Communication Protocols*

| **Protocol** | **Year** | **Key Idea** | **Basis States** | **Security Basis** |
|---|---|---|---|---|
| **BB84**[52] | 1984 | Two conjugate bases | 4 | Heisenberg Uncertainty |
| **B92**[53] | 1992 | Two non-orthogonal states | 2 | State Indistinguishability |
| **E91**[54] | 1991 | Entanglement + Bell test | Entangled | Bell Inequality Violation |
| **SGS04**[55] | 2004 | Two-way qubit reuse | 2 | Ping-pong mechanism |

This section further elucidates the approach for simulating BB84, B92, E91, and SGS04 protocols as given below:

## 3.2.1 BB84

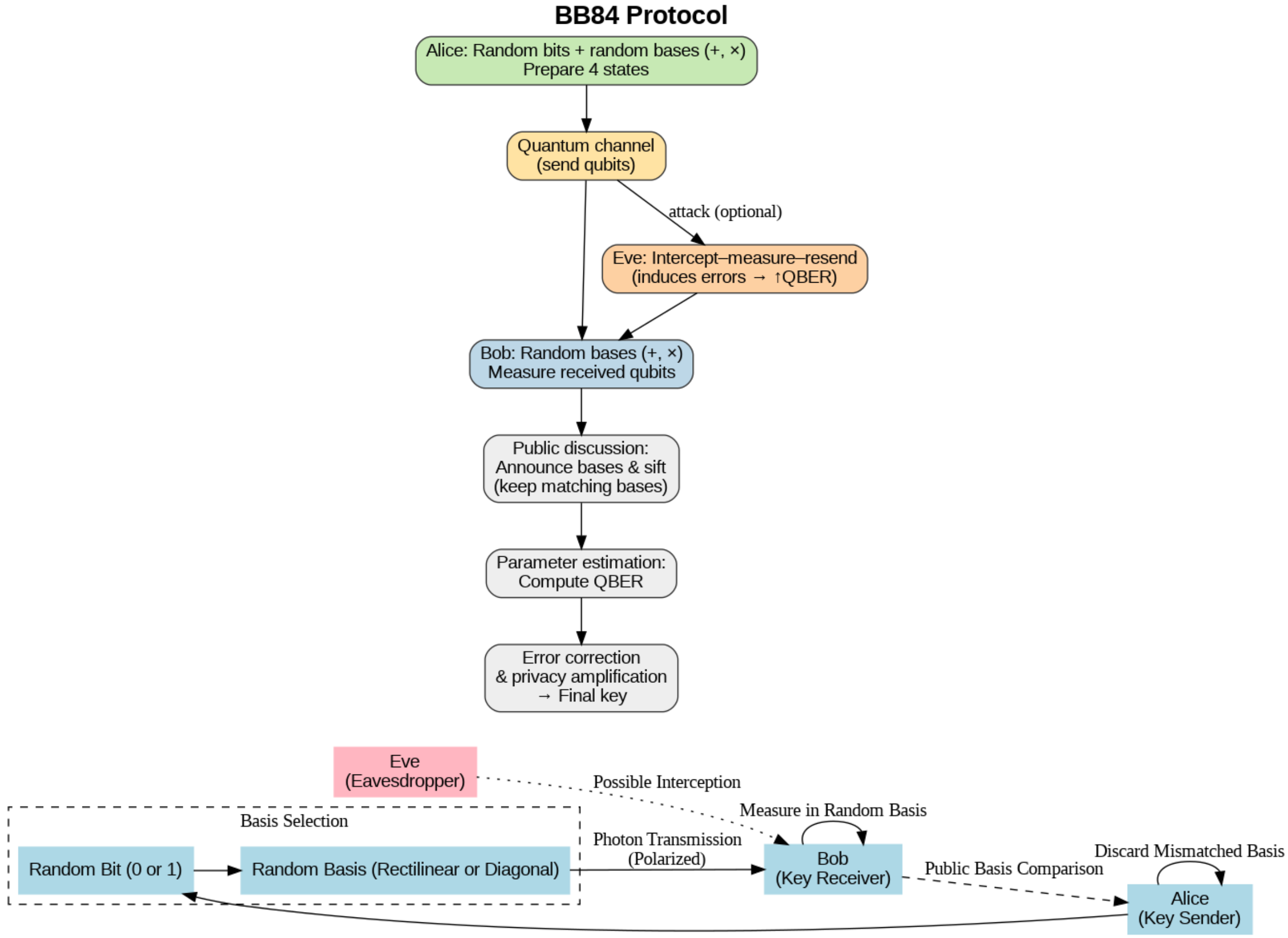

*Fig. 5: BB84 Protocol*

Fig. 5, the Four-State BB84 Protocol, which prepares and measures as Alice uses two mutually unbiased bases (+ and ×) to encode random bits into qubits. Bob uses randomly selected bases to measure each qubit. The final secure key is generated by applying error correction and privacy amplification, estimating QBER to detect eavesdropping, and keeping only matching-based results (sifting) after public basis comparison. Measurement disturbance and the no-cloning theorem are essential to security.

## 3.2.2 B92

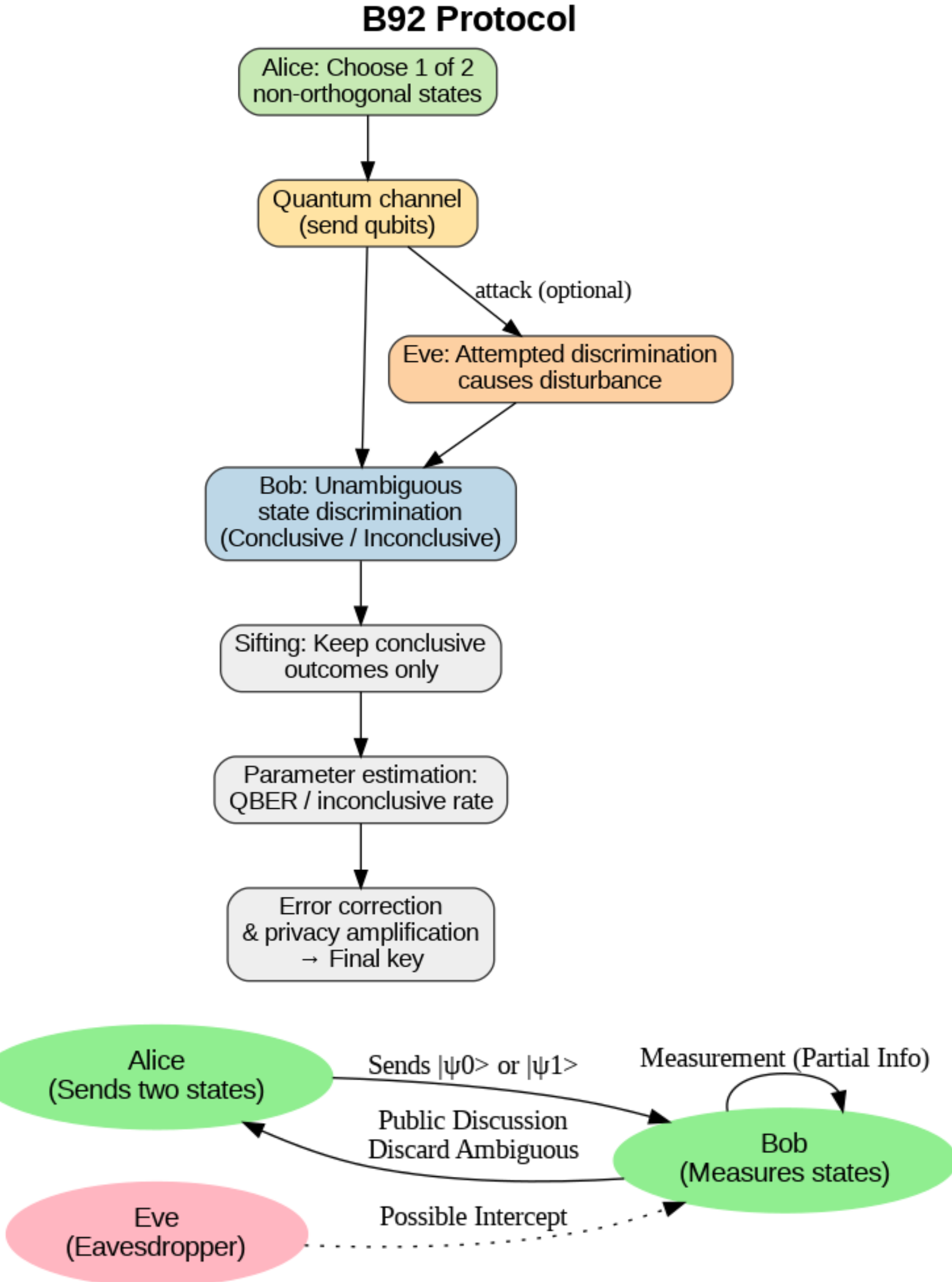

*Fig. 6: B92 Protocol*

Fig. 6, the Two Non-Orthogonal States under the B92 Protocol, uses just two non-orthogonal quantum states and streamlines BB84. Bob does clear state discrimination, yielding results that are either decisive or inconclusive. For key generation, only definitive results are retained. The final key is extracted using traditional post-processing after any eavesdropping effort creates observable disruptions in the error or inconclusive rate.

### 3.2.3 E91

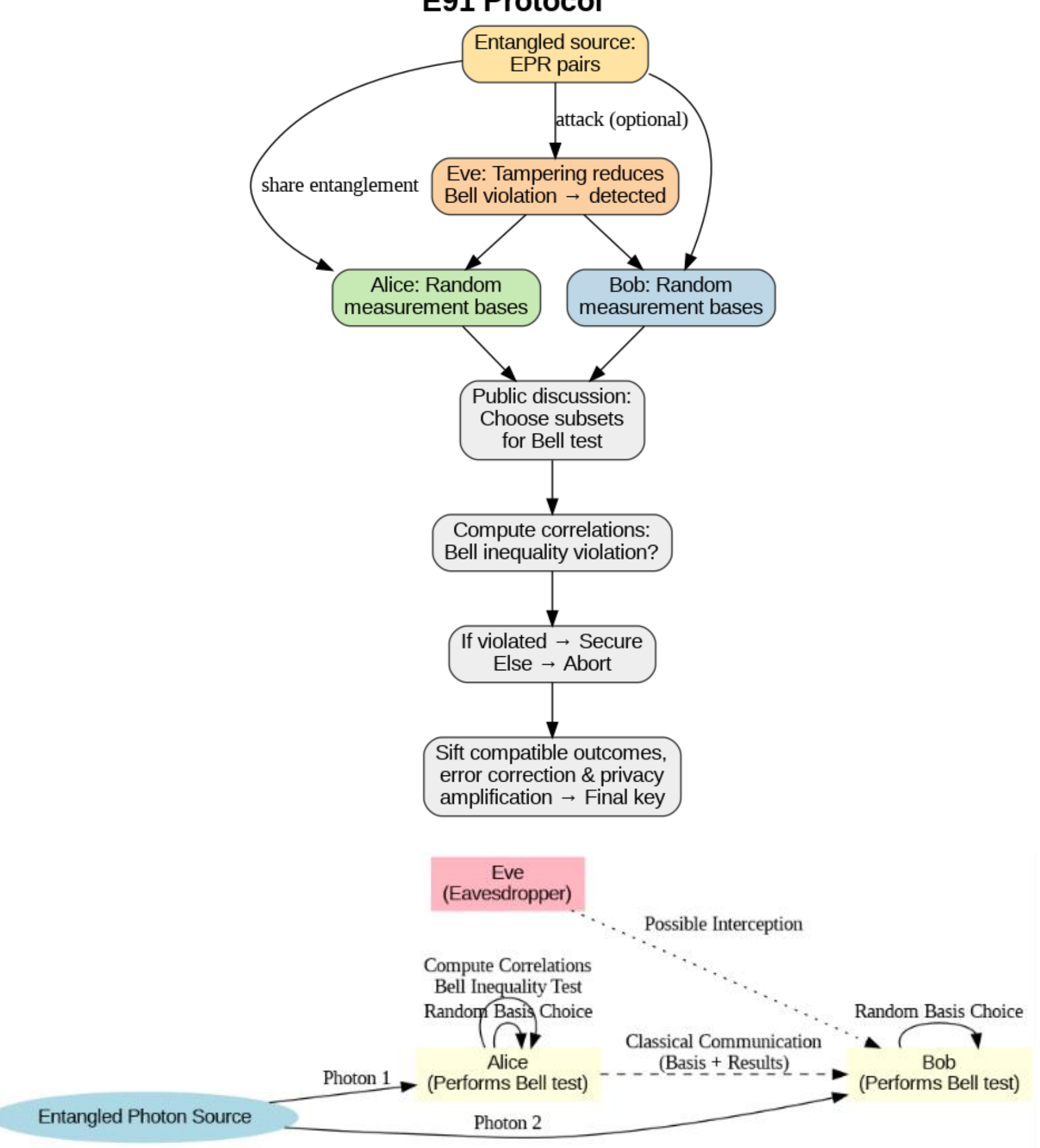

*Fig. 7: E91 Protocol*

Fig. 7, Entanglement-Based E91 Protocol (Bell-Test Security), depicts the method, as Alice and Bob divide up the entangled EPR pairs used by E91. A portion of the data is utilized to test for violations of Bell inequality, and both measures are taken on randomly selected bases. The channel is deemed secure if correlations defy Bell's inequality. The final key is produced by sorting and processing the remaining correlated results. Nonlocal correlations and quantum entanglement are the foundation of security.

### 3.2.4 SGS04

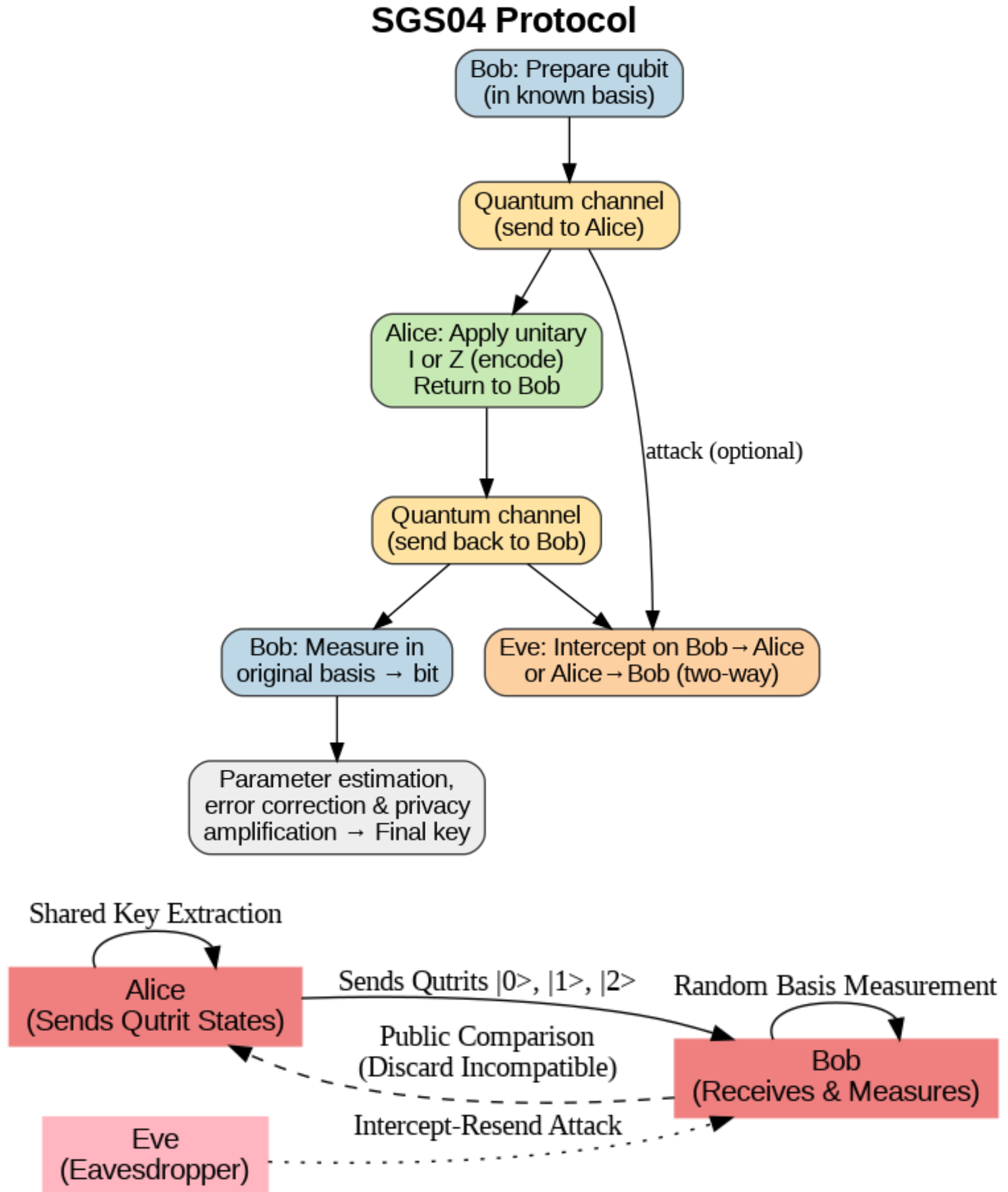

*Fig. 8: SGS04 Protocol*

Fig. 8, the Two-Way Deterministic QKD Protocol (SGS04), illustrates the approach as Bob produces a qubit and transmits it to Alice via the two-way SGS04 protocol. Alice does a unitary operation (I or Z) to encode data before returning it. Bob extracts the bit deterministically by measuring in the original basis. Before the final key is extracted, parameter estimation is essential to identify attacks on either path, since the qubit travels twice.

## 3.3 SCADA datasets in Wireshark PCAP format

In real-world power system operations, for cybersecurity monitoring and in IT-OT conjoined ecosystems, packet-sniffing tools are often deployed. A typical PCAP file generated in Wireshark is shown below:

*Fig. 9: Sample Wireshark PCAP file of SCADA Datasets screenshot*

Fig. 9 shows a standard Wireshark packet-sniffing display pane.

The primary interface is the Packet List Pane (top/main pane), which displays a tabular overview of each collected packet, with one row per packet. Users can effectively filter, sort, and search the data, and it provides a chronological view of network activity. This pane is very helpful for cybersecurity event analysis, fault diagnostics, and traffic flow inspection.

An organised, hierarchical breakdown of the chosen packet is shown in the Packet Details Pane (middle pane). It usually consists of several protocol layers, including the Frame (which records metadata like length and date), Data Link Layer (like Ethernet II or VLAN), Network Layer (IPv4/IPv6), Transport Layer (TCP/UDP), and Application Layer protocols (like HTTP, DNS, Modbus, DNP3, and IEC-61850). It is possible to examine header fields, flags, sequence numbers, checksums, and other protocol-specific characteristics in depth by expanding or contracting each layer.

The raw packet data is shown in both hexadecimal and ASCII formats in the Packet Bytes Pane (bottom pane). Byte offsets are displayed on the left, hexadecimal values are shown in the middle, and ASCII-decoded characters are displayed on the right. This pane's interactivity is one of its main features: selecting any field in the Packet Details Pane instantly highlights the corresponding bytes in the Packet Bytes Pane, enabling accurate low-level examination of packet data.

## 3.4 Approach

This work, which cites previous work where it is demonstrated successfully that using multiple fields of Wireshark SCADA data that are encoded in the BB84 quantum circuits, shows that encryption and decryption have been successful [56] is extended and further broadened in scope, based on four QKD protocols for comparing QBER and thus identifying the best basis for selecting the best protocol in a real telemetry application. Fig. 10 depicts the workflow diagram for the SCADA dataset QKD implementation using quantum circuits for the BB84 protocol, which has been further extended to the other 3 QKD protocols for comparison using the PNNL dataset [57],[58], Day 3: a selected data set. Using Google Colab [59], Qiskit SDK [60] Simulator, the data encoded using the angle encoding technique [61] in corresponding QKD protocol quantum circuits for simulation in Qiskit. The dataset, as mentioned above, contains over 2.5 million records, and the six fields of the PCAP packet data are represented by the corresponding six variables, derived from the PCAP file converted to CSV format. Real quantum hardware is, though, readily available; however, error correction and hardware compatibility with circuit design sensitivity is a growing research areas for large-scale implementation purposes [62].

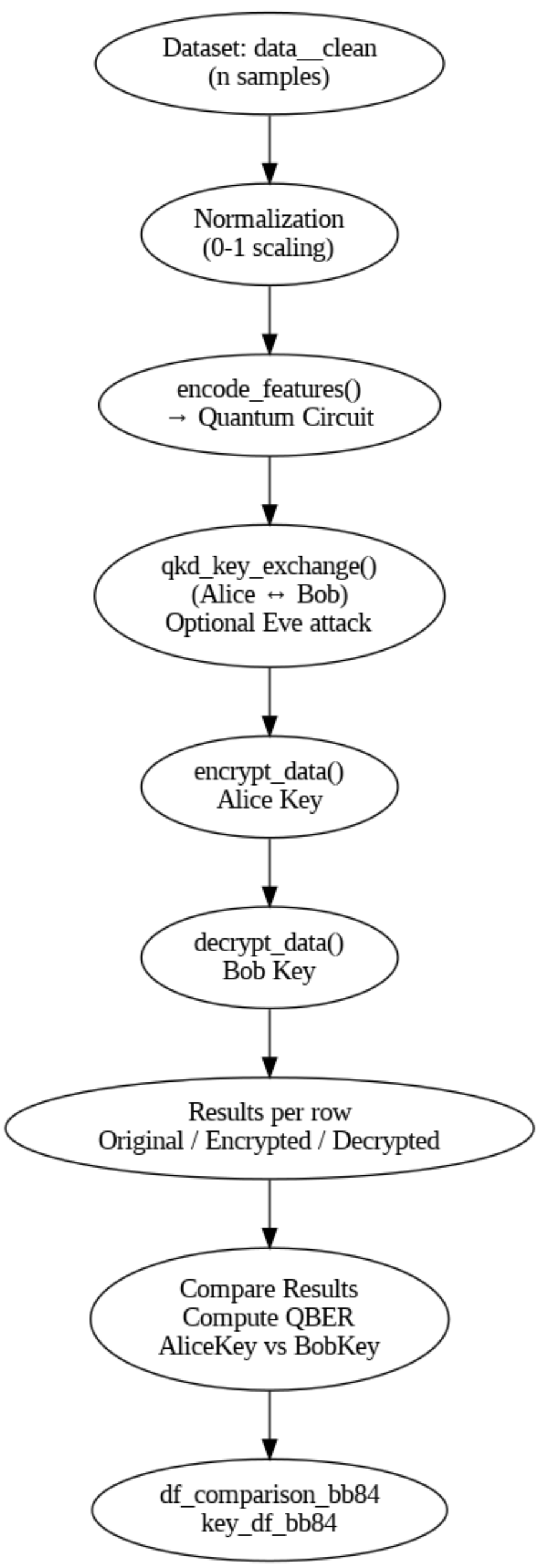

*Fig. 10: Flowchart of the Implementation*

Fig. 10 depicts the entire simulation process of a secure communication system based on QKD and incorporating quantum-feature encoding. After cleaning and normalising the dataset (0–1 scaling), encode_features() encodes classical features into a quantum circuit. Alice and Bob exchange keys using a QKD protocol (qkd_key_exchange()), which can optionally mimic an Eve assault. Bob uses his key to decode the data after Alice uses hers to encrypt it, and the outcomes are compared row by row. Lastly, comparison dataframes for security analysis are generated by evaluating key consistency (AliceKey vs. BobKey) and QBER (Quantum Bit Error Rate). The approach has been iterated for all the four QKD protocols.

The following equations were used in the background to derive the quantum circuits for each protocol, customized to use the six fields of the SCADA data.

Normalization:

$$f_i^{normalized} = \frac{f_i - \min(f)}{\max(f) - \min(f)} \quad (1)$$

*Rotation Angle:*

$$\theta_i = \pi * f_i^{normalized} \quad (2)$$

*Encryption and decryption:*

$$f_i' = \begin{Bmatrix} f_i & if\ k_i = 0 \\ 1 - f_i & if\ k_i = 1 \end{Bmatrix} \quad (3)$$

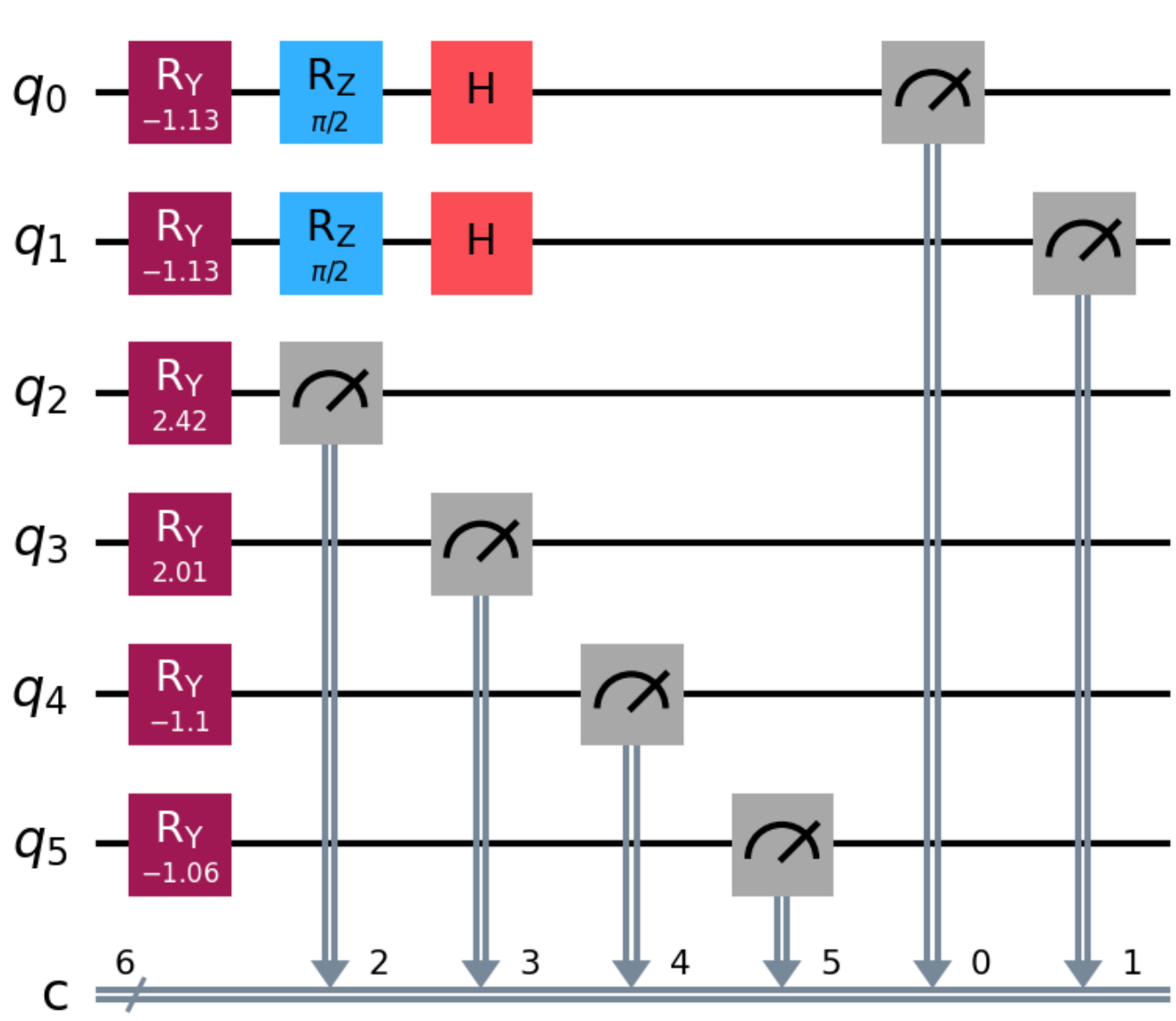

*Fig. 11: A specific Quantum Circuit for the SCADA communication Data for a sample row [0]*

A typical quantum circuit, as shown in Fig. 11, comprises a deep architecture of Hadamard gates, phase flips, and multi-qubit entanglement. The quantum circuit is built using Eq. 4-10.

Encoding stage:

$$\theta_i = \pi * f_i \quad (4)$$

$$R_y(\theta_i) = e^{-i\theta_i \frac{Y}{2}} \quad (5)$$

$$H = \frac{1}{\sqrt{2}} \begin{bmatrix} 1 & 1 \\ 1 & -1 \end{bmatrix} \quad (6)$$

Where the states are:

$$|0\rangle \rightarrow \frac{|0\rangle + |1\rangle}{\sqrt{2}}$$

$$|1\rangle \rightarrow \frac{|0\rangle - |1\rangle}{\sqrt{2}}$$

Controlled-X (CNOT) Gates

Entangling operations:

$$CNOT|c,t\rangle = |c\rangle \otimes X^c|t\rangle \quad (7)$$

Where X is the Pauli X-gate

Pauli Z Gate

$$Z = \begin{bmatrix} 1 & 0 \\ 0 & 1 \end{bmatrix} \quad (8)$$

The layered ansatz

$$U(\vec{\theta}) = \prod_{l=1}^{L} \left[ U_{entangle}{}^{l} \cdot \otimes R(\theta_i{}^{l}) \right] \quad (9)$$

Where L is the number of layers, 4

The final measurement in the Z basis gives the probabilities

$$p(x) = |\langle x | U(\vec{\theta}) | 0^{\otimes n} \rangle|^2 \quad (10)$$

Where x is the bitstring outcome

Fig. 11, a simpler approach for this quantum feature-encoding of SCADA data using single-qubit rotation gates. Each horizontal line represents a qubit, and each qubit is equal to one SCADA variable (six qubits are equal to six variables). The Ry encodes the numerical value of that SCADA variable into the quantum state of each qubit by rotating along the Y-axis of the Bloch sphere when a (θ) gate is applied to each qubit. In practice, the conventional SCADA readings are normalized to a certain range and then mapped to a rotation angle $\theta$ (θ). A small angle retains the qubit close to the $|0\rangle$ state, suggesting a low or nominal value, whereas a larger angle (radians) rotates the qubit closer to $|1\rangle$, indicating a higher or more extreme measurement. Following encoding, the probability of measuring 1 on each qubit instantly reflects the magnitude of the associated SCADA variable. This circuit creates a six-qubit quantum state from six classical SCADA variables. In short, each qubit is a quantum rotation stored as a single SCADA measurement's tiny, probabilistic carrier. To achieve the same outcomes, customized quantum circuits with SCADA six-variable data embedded have been created for the remaining three QKD systems: B92, E91, and SGS04; however, apart from data encoding, the rest of the circuits are tailored to these systems.

# 4.0 Results & Discussion

Based on the above approach in the previous section, the following results surface as given below:

*Table 2: Four protocols —sample 5 rows of QKD comparison*

| | Index | AliceKey | BobKey | KeySize | KeysMatch | QBER |
|---|---|---|---|---|---|---|
| **BB84** | | | | | | |
| **96** | 96 | 00110 | 01110 | 5 | False | 0.20 |
| **98** | 98 | 01 | 01 | 2 | True | 0.00 |
| **5** | 5 | 1001 | 0011 | 4 | False | 0.50 |
| **40** | 40 | 101 | 100 | 3 | False | 0.33 |
| **2** | 2 | 0 | 0 | 1 | True | 0.0 |
| **B92** | | | | | | |
| **99** | 99 | 0101 | 0101 | 4 | True | 0.0 |
| **58** | 58 | 10100 | 10100 | 5 | True | 0.0 |
| **22** | 22 | 1 | 1 | 1 | True | 0.0 |
| **31** | 31 | 10 | 10 | 2 | True | 0.0 |
| **73** | 73 | 000 | 000 | 3 | True | 0.0 |
| **E91** | | | | | | |
| **80** | 80 | 100 | 100 | 3 | True | 0.0 |
| **35** | 35 | 10110 | 10110 | 5 | True | 0.0 |
| **61** | 61 | 111 | 111 | 3 | True | 0.0 |
| **83** | 83 | 10111 | 10111 | 5 | True | 0.0 |
| **4** | 4 | 0 | 0 | 1 | True | 0.0 |
| **SGS04** | | | | | | |
| **60** | 60 | 00 | 10 | 2 | False | 0.50 |
| **65** | 65 | 0 | 0 | 1 | True | 0.00 |
| **29** | 29 | 01 | 01 | 2 | True | 0.0 |
| **36** | 36 | 010 | 011 | 3 | False | 0.33 |
| **5** | 5 | 000000 | 000000 | 6 | True | 0.0 |

Table 2 depicts the analysis outcome of the four QKD protocols, and 5 sample rows corresponding to AliceKey, BobKey, KeySize, KeyMatch, and QBER, respectively, from the 100 rows of the dataset. It summarizes the outcome of multiple quantum key distribution (QKD) key comparison instances between Alice and Bob for the sample five rows of the dataset with the corresponding BB84, B92, E91, and SGS04 technique. For each index, the bit strings generated independently by Alice and Bob are shown along with the resulting key size, and a boolean indicator specifies whether the keys match exactly. When the keys do not match, the Quantum Bit Error Rate (QBER) quantifies the fraction of mismatched bits; higher QBER values (e.g., 0.60–0.67) indicate significant discrepancies likely due to noise or eavesdropping, whereas matching keys yield a QBER of zero, reflecting error-free key agreement.

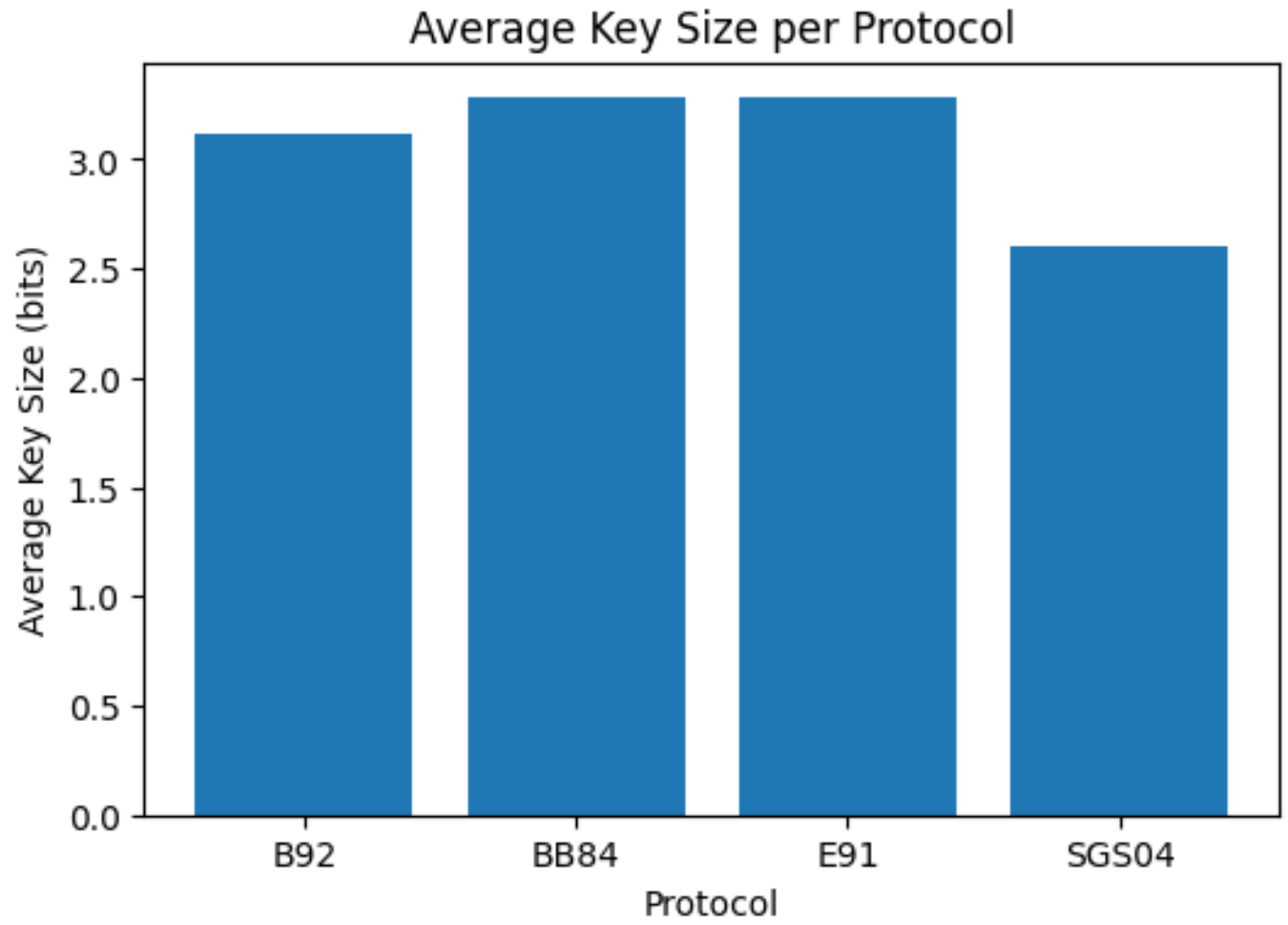

*Fig. 12: Bar plot of Ave Key Size per QKD protocols*

For the first 100 samples of the selected PNNL Day 3 dataset, Fig. 12 shows the average key size for each QKD protocol.

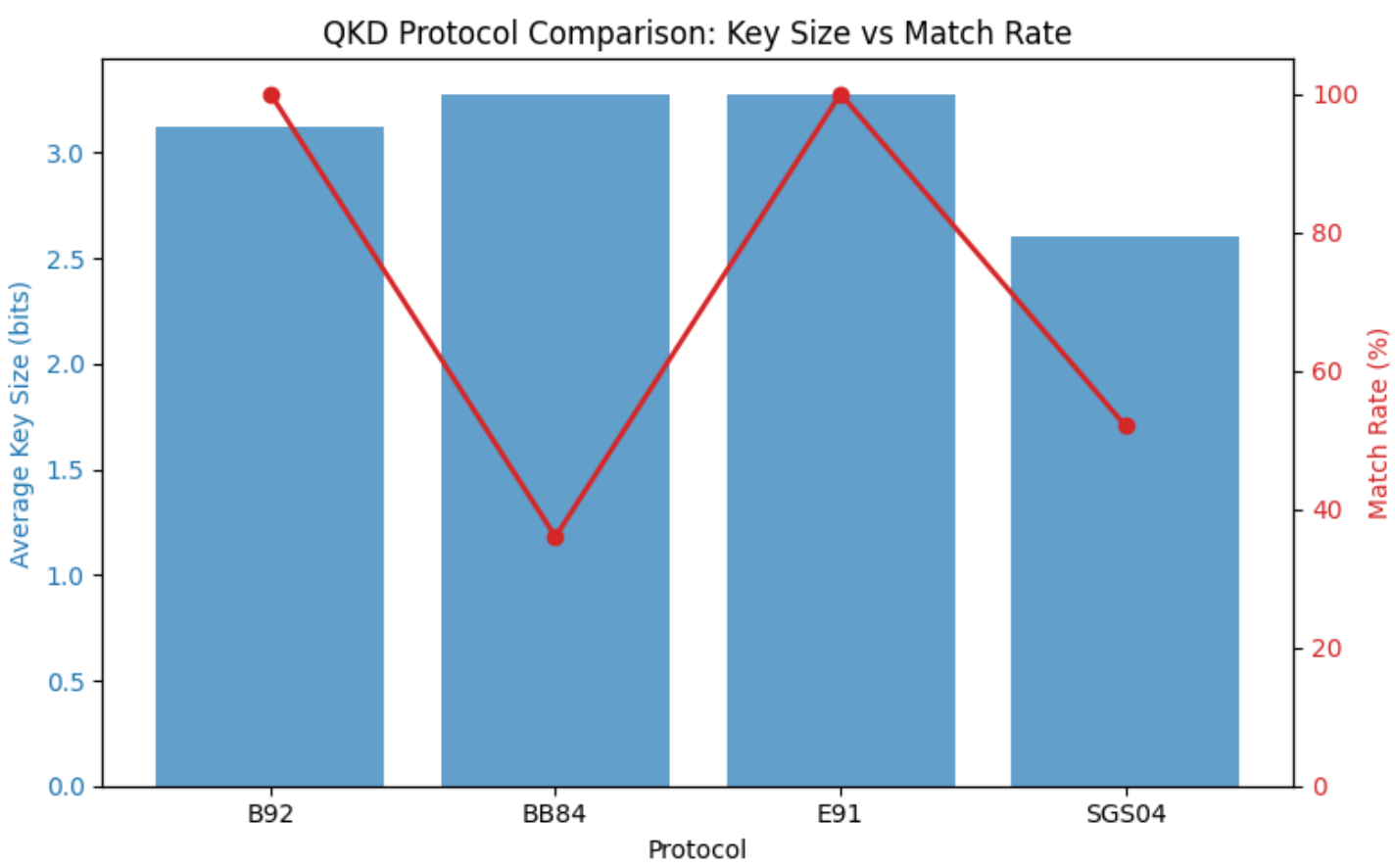

*Fig. 13: Bar Plot of Key Size Vs Match Rate for All the QKD Protocols*

For the first 100 samples of the Day 3 dataset, Fig. 13 shows the average key size vs. match rate percentage for the four QKD protocols.

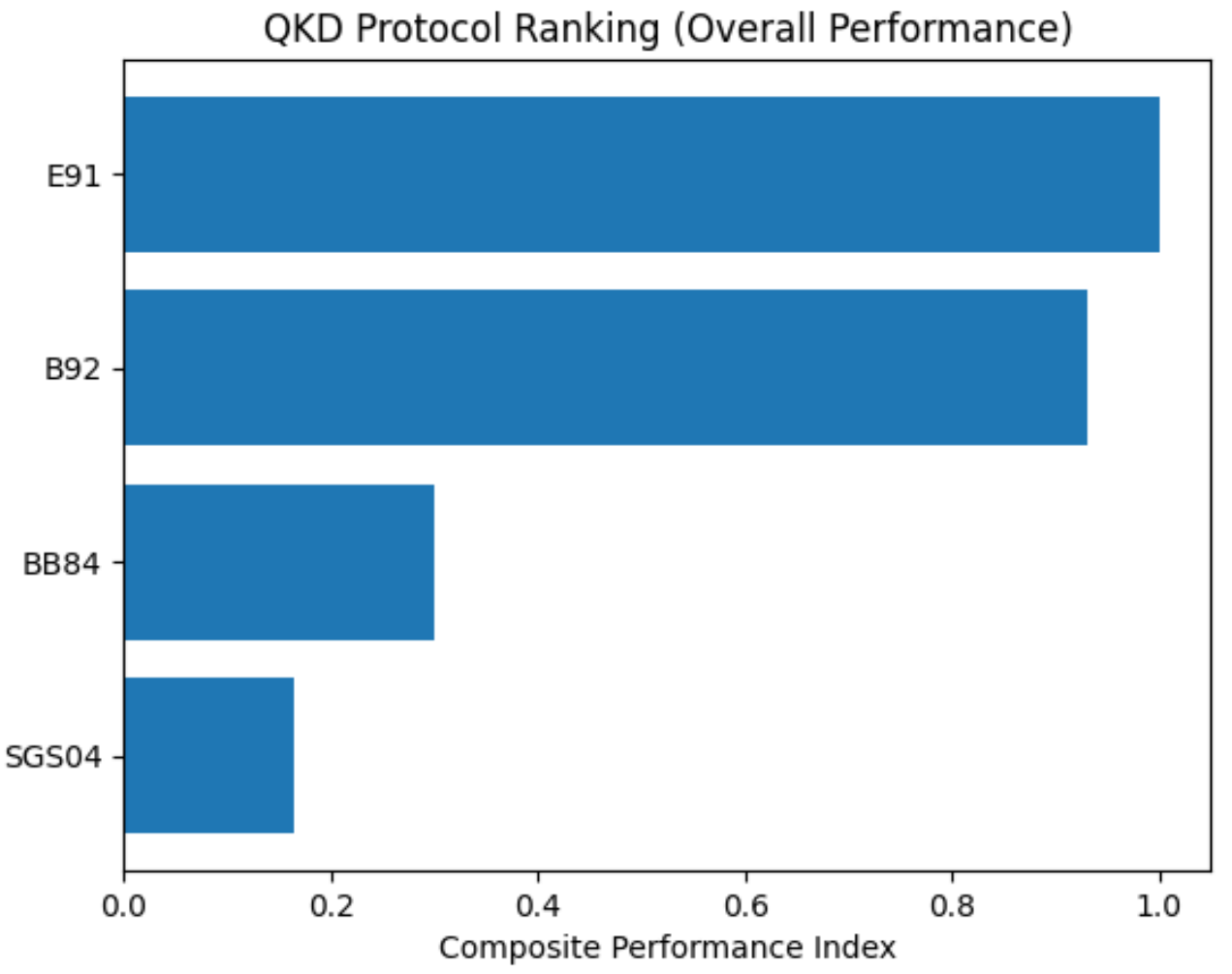

*Fig. 14: Comparison Bar Chart of all the QKD protocols*

Because the third protocol, BB84, achieves a large key size but performs poorly on match rate and error metrics, it is significantly skewed, suggesting that while many bits are generated, a sizable portion are unreliable, which is undesirable for practical SCADA security. In fact, this would result in frequent reconciliations or key rejections because the fourth protocol, SGS04 (red), scores poorly across all aspects, indicating poor key agreement and high inconsistency.

Not all QKD techniques scale in the same way as SCADA traffic characteristics, as the plot illustrates. techniques that balance key length with consistency and low error clearly outperform those optimized solely for raw key generation. A composite performance index plot weights = { "AvgKeySize": 0.30, "MatchRate": 0.30, "AvgMaxDiff": 0.20, "AvgMeanDiff": 0.20}(Fig. 14) shows a bar chart comparing the four QKD protocols using the SCADA data.

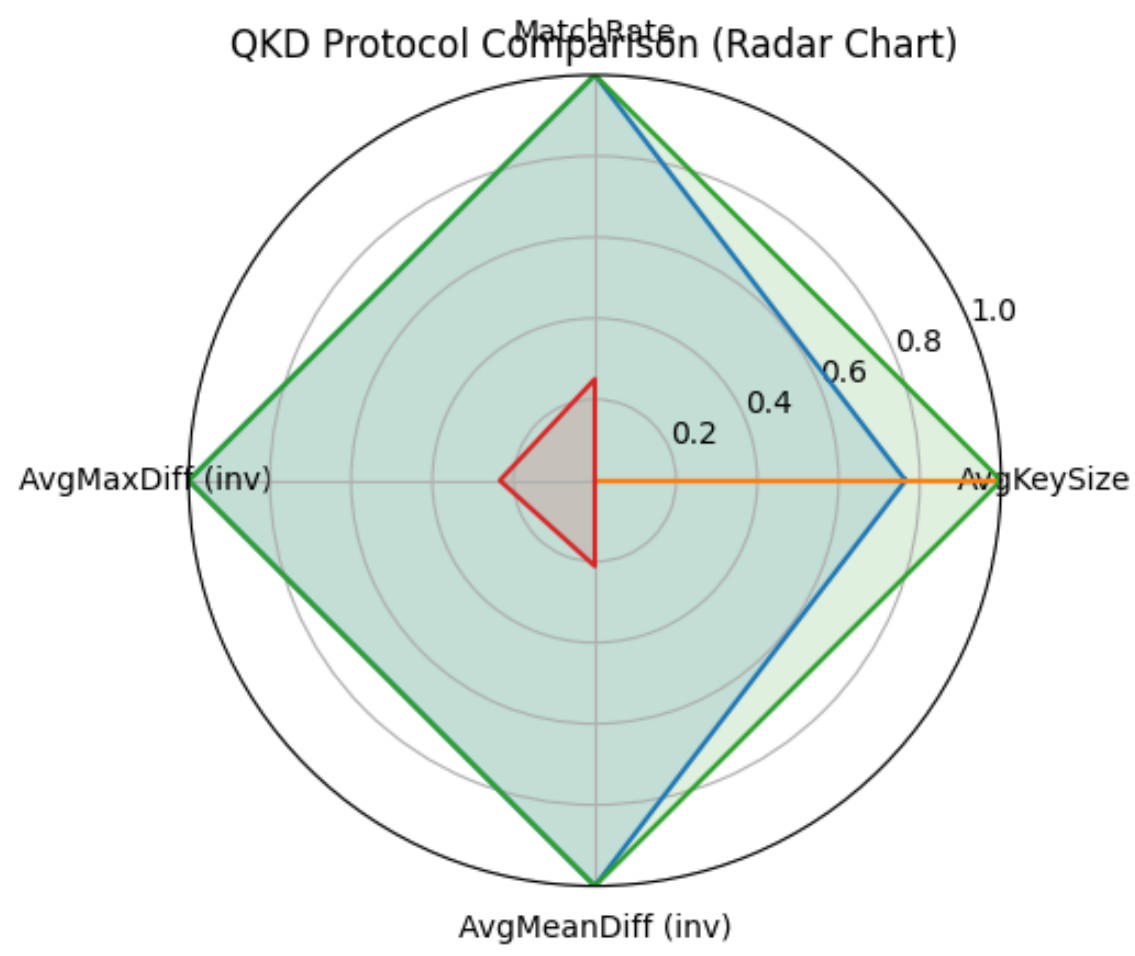

*Fig. 15: Radar Chart comparing all the protocols*

Fig. 15, the radar chart compares QKD techniques; the axes display the normalized performance measures i.e., KeySize, MatchRate, AvgMeanDiff, and AvgMaxDiff, and each color denotes a distinct protocol. Larger, more widely dispersed areas in protocols do better overall across all criteria. This enables rapid visual evaluation of QKD's appropriateness across different criteria, highlighting its inherent trade-offs. Clearly, the E91 protocol stands out among the others. This radar graphic illustrates trade-offs rather than absolute numbers by comparing four QKD algorithms assessed on the same SCADA dataset using four normalised performance criteria.

The best-balanced performance is achieved by a single protocol (the green trace, E91), which consistently reaches the outer boundary on all axes. It achieves the lowest key disagreement (as indicated by the inverted mean and maximum difference metrics), the largest average key size, and perfect or nearly perfect key matching. Because of this, it is the most dependable option for SCADA traffic, where steady, low-error key agreement is essential.

While maintaining a high match rate and minimal disagreement, a second protocol (blue, B92) produces a slightly smaller key size, indicating greater robustness at the expense of lower throughput. Fig. 15 further compares all QKD protocols and ranks E91 as the best overall based on the composite performance index, which corroborates Fig. 14.

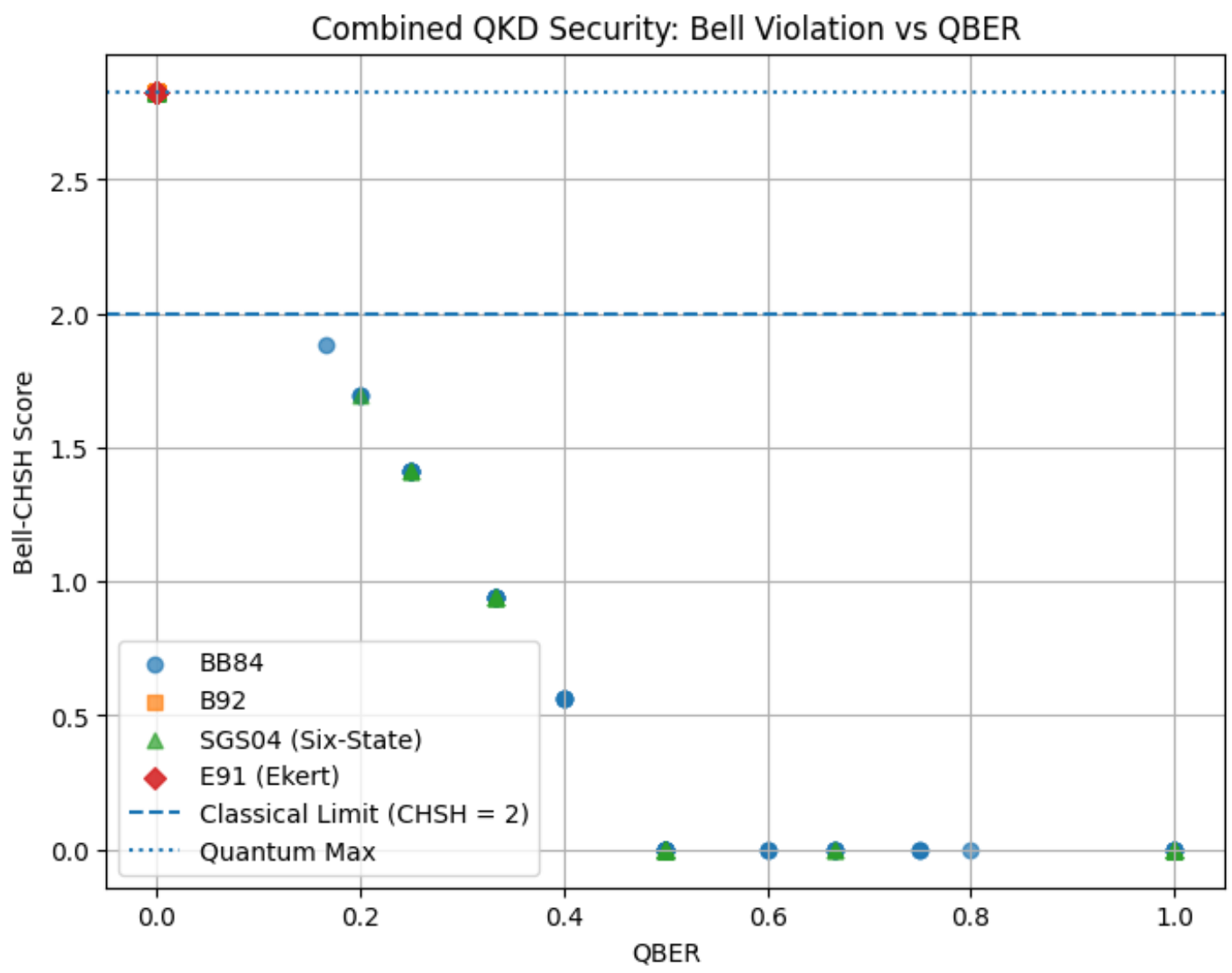

*Fig. 16: Combined security for all QKD – Bell violation vs QBER*

Smart grid communications have been authenticated using quantum secret keys. More precisely, QKD has been used to generate secret keys that facilitate DER communication via the MQTT IoT protocol. The communications between a single photovoltaic (PV) system and a SCADA system constitute the operational idea [8].

Using QKD techniques in the smart grid offers the following advantages: future-proof, immediate, and unconditional security, albeit with certain practical drawbacks. Moreover, system scalability, long-distance end-to-end security, and real-time telemetry are some of the challenges [63]. However, smart grid components range from power networks to smart meters, with interoperability being a key criterion for system operations.

Key sifting is a step in which, once the receiver has measured every received quantum state, the important filtering phase begins. In this step, the receiver initially uses a traditional channel to notify the sender of a list of its measurement bases. In exchange, the sender notifies the recipient of its list of chosen bases via the classical route. Both the sender and the recipient will remove each entry with a different base from their local lists after comparing the received list of bases with the local list. The remaining binary bit sequence is the same on both sides if there is no communication channel fault and no eavesdropper [63].

Through an electric utility's deployment of a trusted node quantum key distribution scheme, it was demonstrated how trusted nodes with practical utility fiber infrastructure may be used to extend the QKD distance constraint, indicating the potential of quantum technologies to provide information security to electric grid equipment [7].

Reportedly, some commercial applications of QKD tends to use the E91 protocol using entanglement [8].

*Table 3: Discrete-variable (DV) QKD comparison*

| **Protocol** | **Year** | **Entanglement?** | **States** | **QKD Class** |
|---|---|---|---|---|
| **BB84** | 1984 | No | 4 discrete | DV-QKD |
| **B92** | 1992 | No | 2 discrete | DV-QKD |
| **E91** | 1991 | Yes | Entangled qubits | DV-QKD |
| **SARG04** | 2004 | No | 4 discrete | DV-QKD |

The CHSH inequality is violated in the entanglement-based E91 protocol to certify security. The Tsirelson bound $S\leq$ 2 2 S≤2 2 [64], which denotes the greatest nonlocal correlations allowed by quantum theory, is the maximum violation that quantum mechanics allows. Degradation of entanglement, possibly because of eavesdropping, is indicated by any decrease in the reported CHSH value S.

The Holevo quantity χ [65], which bounds the maximum amount of classical information that can be extracted from a quantum state, effectively limits the information the adversary can access. Acín et al. (2007) [66] established a device-independent security bound by directly relating Eve's Holevo information to the observed CHSH violation. The Devetak–Winter formula [67] is then used to calculate the resulting secret key rate:

$$K \geq I(A{:}B) - \chi(E{:}B)$$

where $I$ ($A$: $B$) is the mutual information between lawful parties, while Eve's information is constrained by χ(E: B). Therefore, secure key extractability is immediately quantified by the observed Bell violation.

Fig. 16, using CHSH [68], a generalized Bell inequality including correlations between two particles detected under four distinct detector settings is known as the CHSH inequality, which shows that the combined Bell-violation-QBER analysis highlights E91's unique ability to detect eavesdropping via loss of entanglement. At the same time, BB84, B92, and SGS04 rely on error-rate–based disturbance detection with differing robustness.

*Table 4: IEC 62351 mapping scope*

| **IEC 62351 Aspect** | **Code** |
|---|---|
| Secure key establishment | BB84/B84/E91/SGS04 |
| Attack detection | QBER |
| Session keys | AliceKey / BobKey |
| OT traffic protection | Feature encryption |

| Trust decision | QBER threshold |
|---|---|

*Table 5: Real fiber-based conceptual significance*

| QKD **Code Element** | **Real Fiber SCADA Meaning** |
|---|---|
| Qubits | Photons in fiber |
| Z / X basis | Polarization/phase |
| Eve intercept | Fiber tap / MITM |
| Measurement collapse | Detectable disturbance |
| QBER | Fiber security health |

# 5.0 Conclusion

The paper demonstrated that it involves four QKD protocols to encrypt SCADA traffic, validate security via QBER, enable attack-aware SCADA analytics, provide protocol-level trust scoring, support adaptive key acceptance/rejection, and integrate with SCADA or even phasor measurement units (PMU)/wide area monitoring systems (WAMS) links, as applicable.

Across four distinct QKD protocols, key exchange over optical fiber enables intrusion-aware SCADA communication, in which QBER serves as a physical-layer trust metric prior to IEC-62351–compliant encryption of power system telemetry. The results are consistent and encouraging, and the approach further aligns with IEC 62351, Table 4.

Since power system operations worldwide are predominantly fiber-optic-based, the practicality of this approach is supported by the EMI immunity, which ensures stable quantum states; dedicated utility fiber enables operation through a trusted channel; low latency allows frequent key refresh, and high availability ensures continuous QKD operation.

The simulations in this paper assume these fiber advantages, as quantum circuits beyond simulation for actual hardware quantum computer implementation inherently require quantum electronics due to the various underlying intricacies.

Considering fiber-optic quantum communication as a secure, separate channel, it supports the results presented in Table 4. While conventional cybersecurity threats are defined by the confidentiality–integrity–availability (CIA) triad, in power system applications this priority is often reversed to availability–integrity–confidentiality (AIC), where availability is imperative 24×7.

In this context, the use of QKD for multivariate SCADA data enables generic, protocol-independent deployment at a global scale. As power system networks are designated as national critical information infrastructures (CII), this work provides a promising direction for real-world application and deployment, with the option to adopt either QKD, depending on the size and complexity of operations.

Using real quantum hardware and applying quantum machine learning techniques to large datasets to enable simulation of real-world deployment scenarios and comparative analysis of results has been considered a future direction.